\documentclass[sigconf,authorversion,nonacm]{acmart}
\usepackage{hyperref}
\usepackage{subfig}
\usepackage{diagbox}
\usepackage{balance}
\usepackage{graphicx}
\usepackage{textcomp}
\usepackage{xcolor}
\usepackage{caption}
\usepackage{makecell}
\usepackage{booktabs}
\usepackage[most]{tcolorbox}
\usepackage{afterpage}
\usepackage{algpseudocode}
\usepackage{algorithm}
\usepackage{amsfonts}

\AtBeginDocument{%
  }

\newtcolorbox{mybox}{
enhanced,
boxrule=0pt,frame hidden,
borderline west={4pt}{0pt}{green!75!black},
colback=green!10!white,
sharp corners
}

\newtcolorbox{mybox2}{
    enhanced,
    boxrule=0pt, frame hidden,
    borderline west={4pt}{0pt}{blue!75!black},
    colback=blue!10!white, 
    sharp corners
}

\begin{document}

\title{
Exploring Uncore Frequency Scaling for Heterogeneous Computing
}

\author{Zhong Zheng}
\affiliation{
  \institution{University of Illinois Chicago}
  \country{USA}
}
\email{zzheng33@uic.edu}

\author{Seyfal Sultanov}
\affiliation{
  \institution{University of Illinois Chicago}
  \country{USA}
  }
\email{ssulta24@uic.edu}

\author{Michael E. Papka}
\affiliation{
  \institution{Argonne National Laboratory}
  \institution{University of Illinois Chicago}
  \country{USA}
  }
\email{papka@uic.edu}
\author{Zhiling Lan}
\affiliation{
  \institution{University of Illinois Chicago}
  \institution{Argonne National Laboratory}
  \country{USA}
  }
\email{zlan@uic.edu}

\begin{abstract}
High-performance computing (HPC) systems are essential for scientific discovery and engineering innovation. However, their growing power demands pose significant challenges, particularly as systems scale to the exascale level. 
Prior uncore frequency tuning studies have primarily focused on conventional HPC workloads running on homogeneous systems. As HPC advances toward heterogeneous computing, integrating diverse GPU workloads on heterogeneous CPU-GPU systems, it is crucial to revisit and enhance uncore scaling. Our investigation reveals that uncore frequency scales down only when CPU power approaches its TDP (Thermal Design Power) --- an uncommon scenario in GPU-dominant applications ---resulting in unnecessary power waste in modern heterogeneous computing systems.
To address this, we present MAGUS, a user-transparent uncore frequency scaling runtime for heterogeneous computing. 
Effective uncore tuning is inherently complex, requiring dynamic detection of application execution phases that affect uncore utilization. Moreover, any robust strategy must work across a diverse range of applications, each with unique behaviors and resource requirements. Finally, an efficient runtime should introduce minimal overhead.
We incorporate several key techniques in the design of MAGUS, including monitoring and predicting memory throughput, managing frequent phase transitions, and leveraging vendor-supplied power management support.
We evaluate MAGUS using a diverse set of GPU benchmarks and applications across multiple heterogeneous systems with different CPU and GPU architectures.
The experimental results show that MAGUS achieves up to 27\% energy savings and 26\% energy-delay product (EDP) reduction compared to the default settings while maintaining a performance loss below 5\% and an overhead under 1\%. 

\end{abstract}

\maketitle

\section{Introduction}

 High-performance computing (HPC) plays a vital role in advancing numerous research fields by utilizing extensive high-end computational, memory, storage, and network resources to solve complex problems. However, these capabilities often come at the cost of significant power consumption. As HPC systems scale to the exascale level in recent years, energy consumption has become a critical concern. The Department of Energy (DOE) has articulated a 20–30 MW power consumption goal for future exascale computing systems \cite{DOE_goal}. For instance, the recently deployed Aurora supercomputer at Argonne National Laboratory delivers a sustained performance of 1.01 exaFLOPS while consuming 38.7 MW of power \cite{Aurora}. The growing power demands of such high-end HPC systems highlight the urgent need for energy-efficient solutions to conserve energy while maintaining performance. 

Extensive research has addressed energy efficiency in HPC systems using techniques such as dynamic voltage and frequency scaling (DVFS), frequency scaling, and power capping \cite{ramesh2019understanding,walker2018hardware,wallace2016application,bailey2015finding,freeh2005using,ge2007cpu}. 
A typical CPU consists of core and uncore components. The core encompasses the CPU cores, while the uncore includes the Last Level Cache (LLC), Memory Controller (MC), and Quick Path Interconnect (QPI)\cite{intel_uncore, AMD_uncore,ARM_uncore}. 
Several studies have shown that the uncore contributes significantly to overall power consumption \cite{gupta2012forgotten,core_uncore}.  
Although prior efforts have examined dynamic uncore frequency tuning, they have primarily addressed traditional HPC workloads running on homogeneous CPU systems. As HPC environments increasingly adopt heterogeneous CPU-GPU architectures and execute hybrid workloads, particularly emerging applications that primarily rely on GPUs, it becomes essential to investigate uncore scaling in modern heterogeneous HPC computing environments.

Recognizing the potential for energy savings through uncore frequency scaling (UFS), existing studies have explored various approaches. These include leveraging extensive hardware counters to develop complex mathematical performance models \cite{sundriyal2018core} and training offline neural network models \cite{zhang2024fcufs}. Beyond model-based approaches, model-free solutions have also been investigated \cite{gholkar2019uncore, andre2022duf,guermouche2022combining}. UPScavenger \cite{gholkar2019uncore} is a pioneering approach that dynamically adjusts uncore frequency by detecting phase transitions between compute-intensive and memory-intensive regions. It leverages DRAM power and Instructions Per Cycle (IPC) monitoring without relying on complex models.
Despite these advancements, several limitations remain. 
For instance, some approaches identify phase transitions between compute-intensive and memory-intensive phases by tracking multiple hardware counters, which can result in considerable overhead. 
Others rely on extensive offline profiling and model construction, while some require manual modifications to user code.
More importantly, limited research has examined the impact of uncore frequency scaling on GPU workloads, particularly for emerging AI-enabled HPC applications and AI/ML workloads. These gaps highlight the need for a lightweight approach that can effectively adjust uncore frequency based on application execution phases while imposing minimal overhead and requiring no modifications to application code. Moreover, such a design must accommodate a wide variety of HPC workloads, especially emerging AI/ML workloads and AI-enabled science applications that harness GPUs for specific computations.

In this work, we begin by investigating whether uncore frequency is dynamically adjusted when executing GPU workloads or AI-enabled science applications. While vendor-supplied dynamic uncore scaling solutions may exist \cite{DUF-2022}, our study reveals that uncore frequency is only adjusted when CPU power usage approaches the thermal design power (TDP). In GPU-dominant applications, i.e., workloads relying primarily on GPUs rather than CPU cores, CPUs rarely reach TDP, leaving the uncore frequency at its maximum and causing significant power waste. We further conduct a case study to examine the impact of uncore scaling on application performance and energy consumption. Our findings highlight the need for uncore frequency scaling to reduce power waste, hence optimizing energy efficiency in heterogeneous computing environments.

Next, we present MAGUS, a model-free, lightweight, and user-transparent runtime for automated uncore frequency tuning of modern HPC workloads on heterogeneous CPU-GPU systems. The design of MAGUS incorporates two primary techniques. First, it provides a lightweight yet effective method for quickly detecting execution phases that affect uncore utilization. Rather than relying on multiple hardware counters for phase detection, MAGUS adopts a single uncore metric, memory throughput, to reduce runtime overhead. Second, inspired by the work \cite{ding2023dps}, we introduce the concept of \emph{memory dynamics} into MAGUS.  Memory dynamics consists of (a) the first derivative of memory throughput and (b) the frequency of memory throughput changes. 
These features enable MAGUS to predict near-future memory throughput trends and identify frequency memory throughput fluctuations, thereby guiding uncore frequency scaling decisions more effectively. MAGUS is designed as a generic framework, making it  applicable across a wide range of GPU workloads and diverse GPU architectures.

We evaluate MAGUS on three Intel-based systems: (1) two are equipped with Intel Xeon Platinum with Nvidia A100 GPU(s), and (2) the third is equipped with Intel Xeon Max with GPU Max based on the Ponte Vecchio architecture. Intel Xeon Max with GPU Max is the base unit deployed in the Exascale system Aurora \cite{Aurora}.  
We evaluate MAGUS with a suite of representative benchmarks and
applications, including the widely used GPU benchmarks from Altis \cite{hu2020altis, altis-sycl}, ECP proxy applications \cite{ecp}, two real-world HPC applications (LAMMPS and GROMACS), and three ML workloads (UNet, ResNet50, and BERT) from the MLPerf benchmark \cite{farrell2021mlperf}.

We compare MAGUS against two baselines: Intel's default uncore frequency setting and UPS \cite{gholkar2019uncore}, a state-of-the-art uncore frequency scaling method. Our experiments examine application performance, power and energy savings, and runtime overhead across single- and multi-GPU configurations. The results demonstrate that MAGUS delivers up to 27\% energy savings compared to both baseline approaches. Throughout our experiments, it maintains performance degradation below 5\% while introducing only minimal overhead (less than 1\%).

\section{Motivation and Challenges} \label{motivation and challenges}

In modern CPUs and GPUs, the CPU core frequency and GPU streaming multiprocessor (SM) clock speed are dynamically adjusted by the hardware to adapt to workload intensity, thereby avoiding unnecessary energy consumption. 
Uncore components, including the last-level cache, memory controller, and interconnect, consume a significant portion of total power, 5\% to 15\% of total CPU power on average, depending on processor architecture and workload characteristics. For applications less reliant on memory bandwidth or cache performance, power savings can be even greater\cite{DUF-2022}. 
In general, uncore frequency is dynamically tuned only when CPU power approaches the thermal design power (TDP)~\cite{andre2022duf}.

\begin{figure}[htb]
    \centering
        \subfloat[CPU core frequency]{%
        \includegraphics[width=0.3\linewidth]{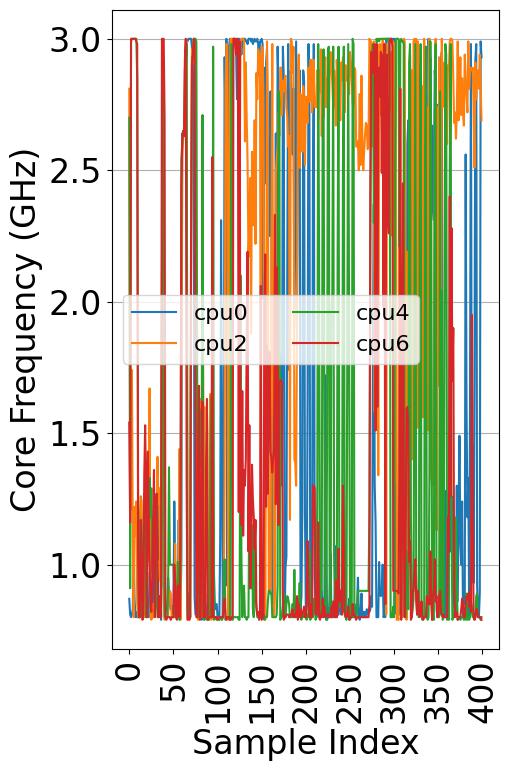}
        \label{fig:cpu_core_freq}
    }
    \subfloat[GPU clock speed]{%
        \includegraphics[width=0.3\linewidth]{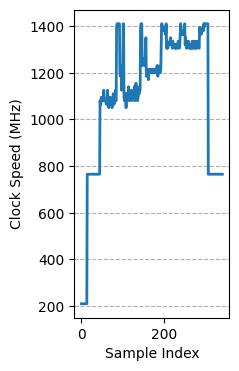}
        \label{fig:gpu_clock}
    }
    \subfloat[Uncore frequency]{%
        \includegraphics[width=0.3\linewidth]{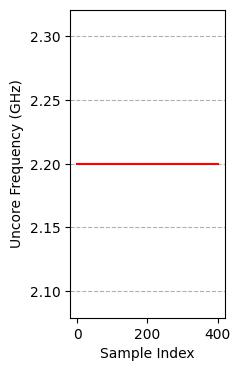}
        \label{fig:uncore_freq}
    }
    \caption{UNet characterization on a heterogeneous Intel Xeon CPU–A100 GPU node. Each socket contains 40 hardware cores; for readability, we plot the core frequency of only four of these cores.}
    \vspace{-5pt}
\end{figure}

To validate this behavior, we analyze a number of applications on a Chameleon system equipped with an Intel Xeon Platinum 8380 CPU and an NVIDIA A100 40GB GPU\cite{keahey2019}. We focus on GPU-dominant workloads --- applications that heavily utilize GPU computational resources while placing minimal to moderate demands on the CPU.
Figure~\ref{fig:cpu_core_freq}-\ref{fig:uncore_freq} presents a case study of UNet, a convolutional neural network for image segmentation \cite{unet}.
As depicted in the figure, CPU core frequency and GPU SM clock speed are tuned dynamically by hardware default setting based on workload demands;
however, the uncore frequency consistently remains at its maximum. 

In the default uncore frequency scaling settings of Intel systems, the uncore frequency is reduced only when the CPU package power approaches the thermal design power (TDP)\cite{andre2022duf}, and we also observe the same phenomenon on other Intel systems including Intel Xeon Platinum and Intel Xeon Max processors. In practice, CPU package power rarely approaches TDP when running GPU-dominant applications, as these workloads are typically not CPU intensive as traditional CPU-only HPC applications.



\begin{figure}[htb]
    \centering
    \subfloat[Max Uncore Frequency (2.2 GHz)]{%
        \includegraphics[width=0.48\linewidth]{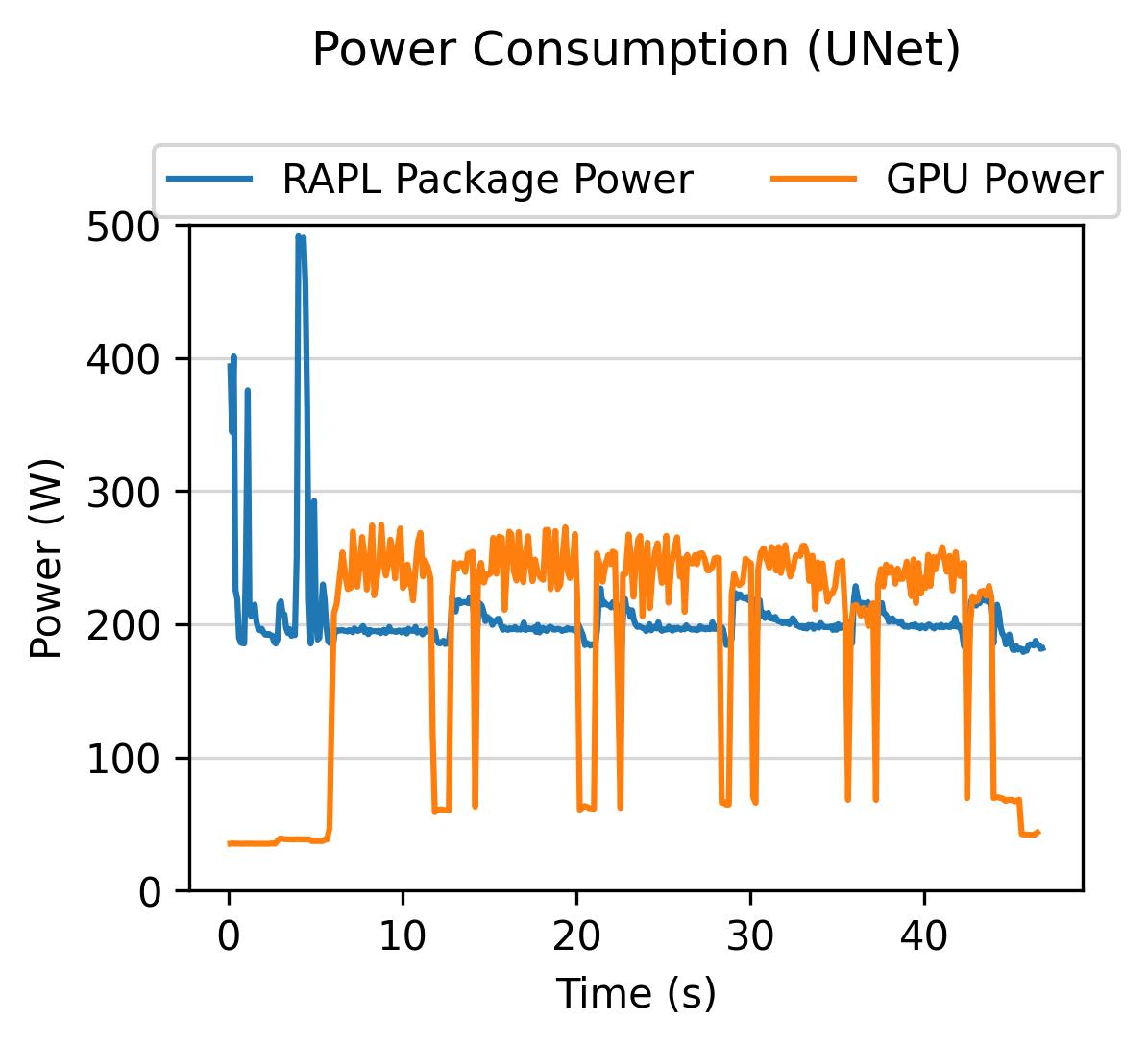}
        \label{fig:UNet_max_uncore}
    }
    \hfill
    \subfloat[Min Uncore Frequency (0.8 GHz)]{%
        \includegraphics[width=0.48\linewidth]{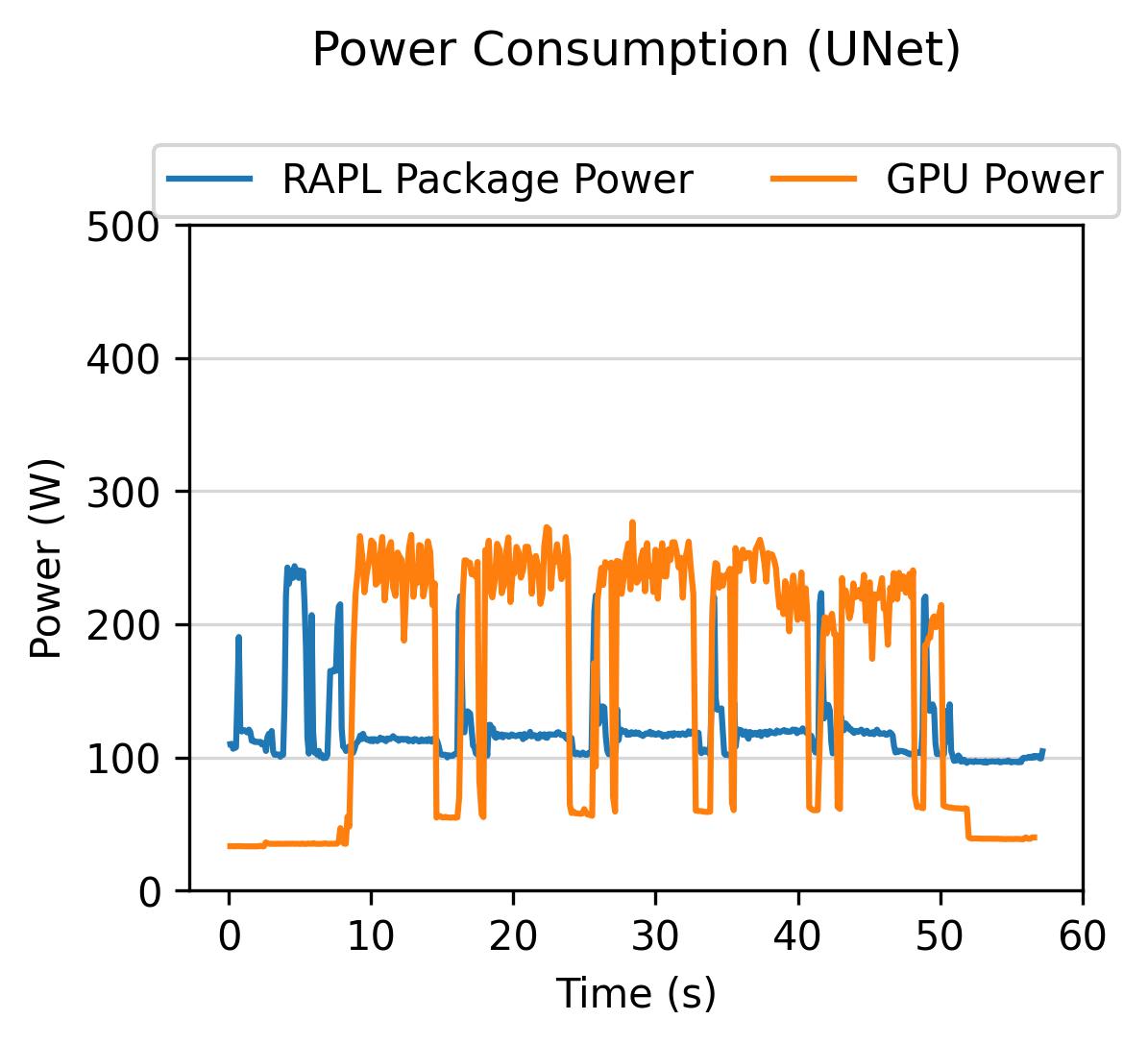}
        \label{fig:UNet_min_uncore}
    }
    \caption{Power profiles of UNet training under different uncore frequencies: max (2.2 GHz) versus min (0.8 GHz). 
    }
    \label{fig:Unet_uncore}
    \vspace{-5pt}
\end{figure}

To analyze the relationship between uncore frequency, power consumption, and performance, we conduct a case study of UNet under various uncore settings (Figure \ref{fig:Unet_uncore}). Using a reduced dataset to maintain the iterative nature of deep neural network training, we compared two scenarios: maximum (2.2 GHz) and minimum (0.8 GHz) uncore frequencies. Setting the uncore frequency to its minimum value reduced CPU package power consumption by 42\% and total energy consumption by 13\%. However, this aggressive power reduction increased execution time by 23\%, highlighting the critical trade-off between energy savings and performance. This analysis shows that reducing the uncore frequency effectively lowers power and energy consumption. However, setting the uncore frequency to its minimum without considering application demands can significantly degrade performance, especially when high memory bandwidth is required. 
During neural network training, distinct power consumption spikes occur during two critical operations: initial data loading and preprocessing and model parameter updates. These observations suggest that we need a dynamic approach to scale uncore frequency based on workload demands, rather than maintaining a constant minimum or maximum frequency.

Previous studies have explored uncore frequency scaling for power efficiency. A notable example is the UPS method proposed by Gholkar et al. \cite{gholkar2019uncore}, which dynamically adjusts uncore frequency by detecting phase transitions between computation-intensive and memory-intensive operations based on DRAM power and IPC changes. However, UPS primarily focuses on traditional CPU-only applications and systems. 
Over the past few years, HPC has undergone significant evolution, with GPU workloads becoming increasingly prevalent and heterogeneous CPU-GPU systems widely adopted. It is therefore time to revisit uncore frequency scaling and understand its implications for emerging GPU workloads in modern heterogeneous environments.
Additionally, UPS relies on active monitoring of IPC, which requires reading instructions retired and CPU cycles through MSRs (Model-Specific Registers) for each core, introducing considerable runtime overhead (details in Section \ref{overhead analysis}). 

The above limitations motivate us to develop MAGUS, an uncore frequency scaling runtime specifically designed for emerging heterogeneous computing environments.

\textbf{Challenges.}
Uncore frequency tuning is inherently complex, as it requires dynamic detection of execution phases that affect uncore utilization. Applications often exhibit multiple power phases with varying frequencies of change, making it crucial to determine how much to increase or decrease uncore frequency at each adjustment to balance energy savings with performance. A robust strategy must not only accommodate a wide range of applications, each with unique behaviors and resource requirements, but also incur minimal runtime overhead. Specifically, we identify the following key challenges:

\begin{enumerate}
\item \emph{Heterogeneity in GPU workloads.} 
GPU workloads frequently alternate between memory accesses and GPU computations at fine-grained intervals. They also differ significantly in their memory access and computation patterns, especially when comparing scientific simulations with machine learning models. Developing a uniform phase detection mechanism that effectively adapts to such diverse behaviors is difficult, requiring tailored strategies to accommodate the unique behaviors of each application.


\item \emph{Selection of uncore metrics.} Modern processors provide multiple uncore metrics, such as IPC, DRAM power, CPI, flops, LLC\_misses, that can guide uncore frequency scaling. 
However, monitoring numerous hardware counters can introduce substantial runtime overhead, particularly in systems with large core counts, where accessing Model-Specific Registers (MSRs) at scale becomes resource-intensive. 
For instance, active IPC monitoring requires accessing Model-Specific Registers (MSRs) to read instructions retired and CPU cycles for each core, a process that becomes increasingly resource-intensive as CPU core counts scale. 
Identifying a minimal yet sufficiently informative metric is critical to achieving both low-overhead data collection and effective tuning decisions.


\item \emph{Frequent phase changes.} In GPU workloads, the transition between compute- and memory-intensive phases can occur at microsecond or millisecond timescales. Capturing these rapid transitions and responding to them without adding significant overhead is challenging.

\end{enumerate}


\section{MAGUS Design} \label{design}

In this section we describe the design of MAGUS, a memory-throughput-based uncore frequency scaling runtime for heterogeneous computing. MAGUS leverages Intel's Performance Counter Monitor (PCM) API \cite{PCM} for monitoring system metrics, and Model-Specific Registers (MSRs) for hardware control. While currently implemented using these Intel-specific interfaces, MAGUS's design principles can be adapted to other processors, such as AMD and ARM, provided they offer interfaces for reading memory throughput data and enabling uncore frequency scaling.

To address the challenges listed in \S2, we adopt \emph{memory throughput} as our primary metric because it directly reflects the activity intensity of uncore components, making it a reliable and effective indicator. Monitoring memory throughput over time provides critical insights for dynamically adjusting uncore frequency, thereby improving energy efficiency. Furthermore, unlike reading IPC data via MSRs from each CPU core, obtaining memory throughput data through Intel's PCM API introduces significantly lower runtime overhead, thereby improving the overall efficiency of the process.

\begin{figure}
    \centering
    \includegraphics[width=1\linewidth]{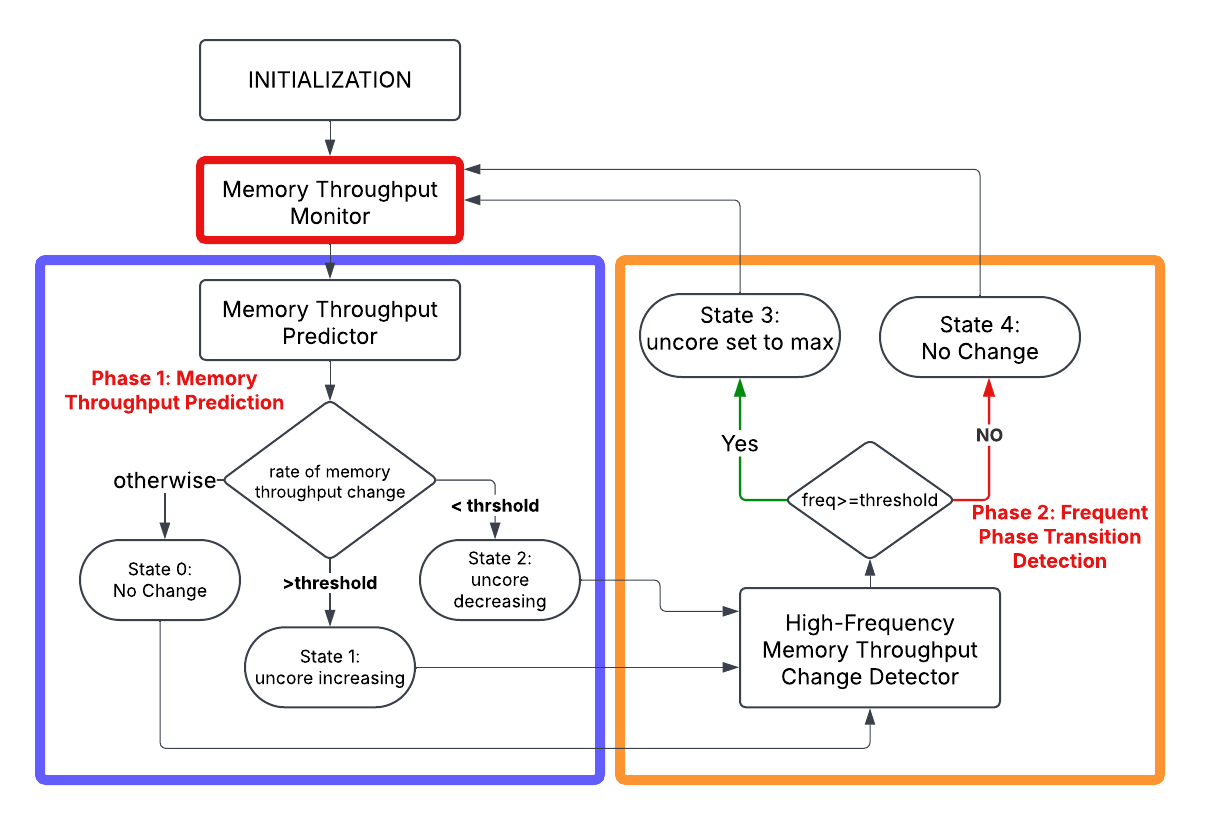}
    \caption{MAGUS Overview. MAGUS comprises three main components: (1) Memory Throughput Monitor, (2) Memory Throughput Predictor, and (3) High-Frequency Memory Throughput Changing Detector, each being highlighted in a different color.}
    \label{fig:magus_control_logis}
    \vspace{-15pt}
\end{figure}

MAGUS operates through two key phases, \emph{memory throughput prediction} and \emph{frequent phase transition detection}, to overcome the major hurdles identified earlier. 
First, rather than merely detecting phase transitions between memory-intensive and compute-intensive phases, MAGUS employs the concept of \emph{memory dynamics}. Specifically, MAGUS uses the first derivative of memory throughput data to anticipate near-future trends as shown in Algorithm 1. This simple yet effective approach allows the runtime to adapt to a broad spectrum of workload behaviors without the overhead of multiple hardware counters. 
Second, MAGUS incorporates a lightweight algorithm listed in Algorithm 2 to automatically detect high-frequency changes in memory throughput, ensuring timely adjustments to uncore frequency and addressing the challenge of high-frequency phase changes.
Figure \ref{fig:magus_control_logis} presents the MAGUS flowchart, and the following subsections describe each phase in detail.

\subsection{Phase 1: Predicting Memory Throughput} \label{predict}

Memory throughput can fluctuate dramatically, either increasing or decreasing sharply. These fluctuations indicate that while applications periodically require maximum uncore frequency for optimal performance, maintaining such high frequency is often unnecessary during less intensive periods. Anticipating these shifts allows for proactive uncore frequency adjustments to match upcoming memory throughput demands. Offline performance profiling methods are impractical, as each application exhibits unique memory access patterns, and previously unseen applications introduce further variability. 
To address this, we introduce the concept of \emph{memory dynamics}, which includes both \emph{the first derivative} and \emph{the frequency of changes} in memory throughput. Specifically, we leverage the first derivative of memory throughput over short intervals to anticipate near-future demands. This predictive approach enables timely and efficient uncore frequency scaling, ensuring responsiveness to workload variations while maintaining energy efficiency. The frequency component will be discussed in Phase 2.

The detailed procedure is outlined in Algorithm 1. We maintain a fixed-size first-in first-out queue to record memory throughput history. The function MEM\_throughput\_TREND\_PREDICTION is invoked periodically to compute the first derivative of memory throughput over a specified interval. If the first derivative exceeds the inc\_threshold, it indicates that memory throughput is likely to increase sharply in the near future. Conversely, if the derivative falls below the dec\_threshold, it suggests that memory throughput demands are expected to decrease significantly. The function returns 1 or -1 to signal the runtime to either increase or decrease the uncore frequency accordingly. If the derivative falls between these thresholds, the uncore frequency remains unchanged, ensuring stability and avoiding unnecessary adjustments. This adaptive mechanism allows for timely, predictive adjustments to uncore frequency, balancing energy efficiency with application performance.

The uncore frequency scaling decision generated by the prediction phase is initially recorded as a temporary decision and is not executed immediately. Depending on the outcome of the frequent phase change detection phase (Section \ref{detection}), this decision may be overwritten or left unchanged.

\begin{algorithm}
\caption{Memory Throughput Prediction}
\begin{algorithmic}[1]
\Function{Mem\_throughput\_Trend\_Prediction}{
    $inc\_threshold$,  // Threshold for increasing frequency
    $dec\_threshold$,  // Threshold for decreasing frequency
    $mem\_throughput\_ls$,  // List of memory throughput values
    $direv\_length$    // Derivative calculation window
}

    \State $derivative \gets \frac{\text{mem\_throughput\_ls}[-1] - \text{mem\_throughput\_ls}[0]}{\text{direv\_length}}$

    \If{$derivative > inc\_threshold$}
        \State \Return 1

    \ElsIf{$derivative < dec\_threshold$}
        \State \Return -1

    \Else
        \State \Return 0
    \EndIf
\EndFunction
\end{algorithmic}
\end{algorithm}

\begin{algorithm}
\caption{High Frequency Detection}
\begin{algorithmic}[1]
\Function{high\_freq\_detection}{$high\_freq\_threshold,$ \newline $uncore\_tune\_ls$} 
    \State // uncore\_tune\_ls: indicates whether a uncore frequency tuning incident is triggered for each monitoring step by either binary flag (0 or 1)
    \State $freq \gets \frac{\text{sum}(uncore\_tune\_ls)}{\text{len}(uncore\_tune\_ls)}$
    \If{$freq \geq high\_freq\_threshold$}
        \State \Return True
    \Else
        \State \Return False
    \EndIf
\EndFunction
\end{algorithmic}
\end{algorithm}

\subsection{Phase 2: Detecting Frequent Phase Changes} \label{detection}
Workloads can experience fluctuating and dramatic memory throughput changes over short time intervals, leading to potential frequent uncore frequency scaling. Frequent uncore frequency scaling in this scenario can degrade application performance for two key reasons. First, frequent uncore frequency scaling incurs excessive MSR accesses, adding overhead. Second, when memory throughput fluctuates rapidly and substantially, neither software nor hardware can fully adapt in real time, limiting the system’s ability to meet instantaneous throughput demands. 

To address frequent memory throughput changes, we developed a simple yet effective detection algorithm (Algorithm 2) that identifies periods of high-frequency fluctuations. During these periods, MAGUS maintains maximum uncore frequency to ensure consistent access to maximum memory bandwidth, preventing performance degradation from constant frequency adjustments. We maintain a first-in-first-out queue called uncore\_tune\_ls, which uses a binary flag (0 or 1) to record whether a potential uncore frequency scaling event should occur based on the uncore frequency scaling decision made by the prediction phase (see Algorithm 1). If the rate of triggered UFS events (either an increase or decrease in uncore frequency) exceeds a threshold of 0.6, it indicates that memory throughput is fluctuating frequently, which we classify as a high-frequency status. For example, if the uncore frequency is adjusted more than six times per second, the system is considered to be in a high-frequency memory change state. When this status is detected, MAGUS overrides the temporary decision from the prediction phase and sets the uncore frequency to its maximum to stabilize performance. Otherwise, MAGUS retains and executes the prediction phase decision. Even if the application remains in a high-frequency state, MAGUS continues running the prediction phase in each subsequent decision round, using it to predict memory throughput and log any potential uncore frequency scaling events. Although these logged events are not executed while in a high-frequency state, they inform high-frequency detection in later rounds.

In summary, this detection algorithm ensures that applications experiencing frequent memory throughput fluctuations consistently receive maximum memory bandwidth, mitigating performance degradation during periods of high variability. 

\section{MAGUS Implementation} \label{implementation}
We implement MAGUS in C++ as a user-transparent runtime.  
The default uncore frequencies of compute nodes are set to their minimum values to conserve power when the nodes are idle. Upon the arrival of an application, MAGUS periodically monitors memory throughput using Intel's PCM API \cite{PCM} and  dynamically adjusts uncore frequency based on real-time workload analysis.

\section{Experimental Configuration} \label{config}

\subsection{Heterogeneous Systems} \label{system}

We evaluate MAGUS across three distinct heterogeneous computing systems:
\begin{itemize}
    \item \emph{Intel+A100}: A Chameleon Cloud \cite{keahey2020lessons} system featuring two Intel(R) Xeon(R) Platinum 8380 processors paired with a single NVIDIA A100-40GB GPU. The system supports uncore frequencies from 0.8 GHz to 2.2 GHz and runs Ubuntu 22.04 with CUDA 12.6.
    \item \emph{Intel+4A100}:  It has the same architecture and software environment as the first, except it is equipped with four NVIDIA A100-80GB GPUs interconnected via PCIe.
    \item \emph{Intel+Max1550}: It features the Intel(R) Xeon(R) CPU Max 9462, a Sapphire Rapids architecture processor comprising four compute tiles Intel(R) Data Center GPU Max 1550 based on the Ponte Vecchio architecture and featuring 128 GB of HBM2e memory. The system supports an uncore frequency range of 0.8 GHz to 2.5 GHz, and operates on Ubuntu 24.04.1 LTS with oneAPI DPC++/C++ Compiler 2024.2.1. Intel Xeon Max with GPU Max is the base unit deployed in the Exascale system Aurora \cite{Aurora}.
\end{itemize}

\subsection{Heterogeneous Benchmarks/Applications} \label{benchmarks}
We select a representative suite of HPC benchmarks and applications in our experiments:

\begin{itemize}
\item \textbf{GPU benchmark suite Altis} includes fundamental parallel algorithms widely used in parallel computing and real-world applications \cite{hu2020altis}. Specifically, we use 14 benchmarks from Level 1 and Level 2, excluding Level 0 benchmarks due to short execution times.

\item \textbf{ECP proxy applications} include miniGAN, CRADL, Laghos, and SW4lite \cite{ecp}. miniGAN is a GAN-based benchmark for deep learning in HPC. CRADL integrates adaptive learning with computational science for surrogate modeling. Laghos is a high-order Lagrangian hydrodynamics solver for gas dynamics. SW4lite is a lightweight seismic wave propagation solver for earthquake simulations.

\item \textbf{AI-enabled applications.} AI-enabled applications, where advanced machine learning methods are used for surrogate models or in situ data analysis, is quickly being adopted in science and engineering for tackling complex computational problems. In this study, we use two  AI-enabled applications, GROMACS \cite{van2005gromacs} and LAMMPS \cite{atomic2013lammps}, both being widely used open-source molecular dynamics (MD) simulation packages. 

\item \textbf{MLPerf benchmarks.} Given the increasing adoption of deep neural network training in HPC, we also select ResNet50, UNet, and BERT from the MLPerf benchmark suite \cite{farrell2021mlperf} to represent deep learning workloads. 
\end{itemize}

For both Intel+A100 and Intel+4A100, the applications employ CUDA for GPU computation. In contrast, for the Intel+Max1550 system, we utilize 11 applications from the Altis-SYCL benchmark suite \cite{altis-sycl}, excluding those that cannot be compiled. The other applications mentioned above currently do not have SYCL versions available.

\subsection{Comparison Methods}\label{comparison method}
In our evaluation, we compare MAGUS to two existing methods: 
\begin{itemize}
    \item \textbf{Default uncore frequency scaling (baseline).} In the default settings of Intel Xeon Platinum 8380 and Intel Xeon CPU Max 9462 processors used in this study, the uncore frequency is reduced only when CPU power (package and DRAM) approaches the Thermal Design Power (TDP). However, in our experiments, CPU power rarely reaches TDP, causing the uncore frequency to remain at its maximum.
    \item \textbf{Uncore Power Scavenger (UPS).} The Uncore Power Scavenger is a model-free runtime approach that dynamically adjusts the uncore frequency based on changes in DRAM power and instructions per cycle (IPC) for HPC applications \cite{gholkar2019uncore}. 
    Since we did not find an open-source UPS implementation,
    we implemented UPS based on the methodology described in the paper.
\end{itemize}

\begin{figure*}
    \centering
    \includegraphics[width=0.8\linewidth]{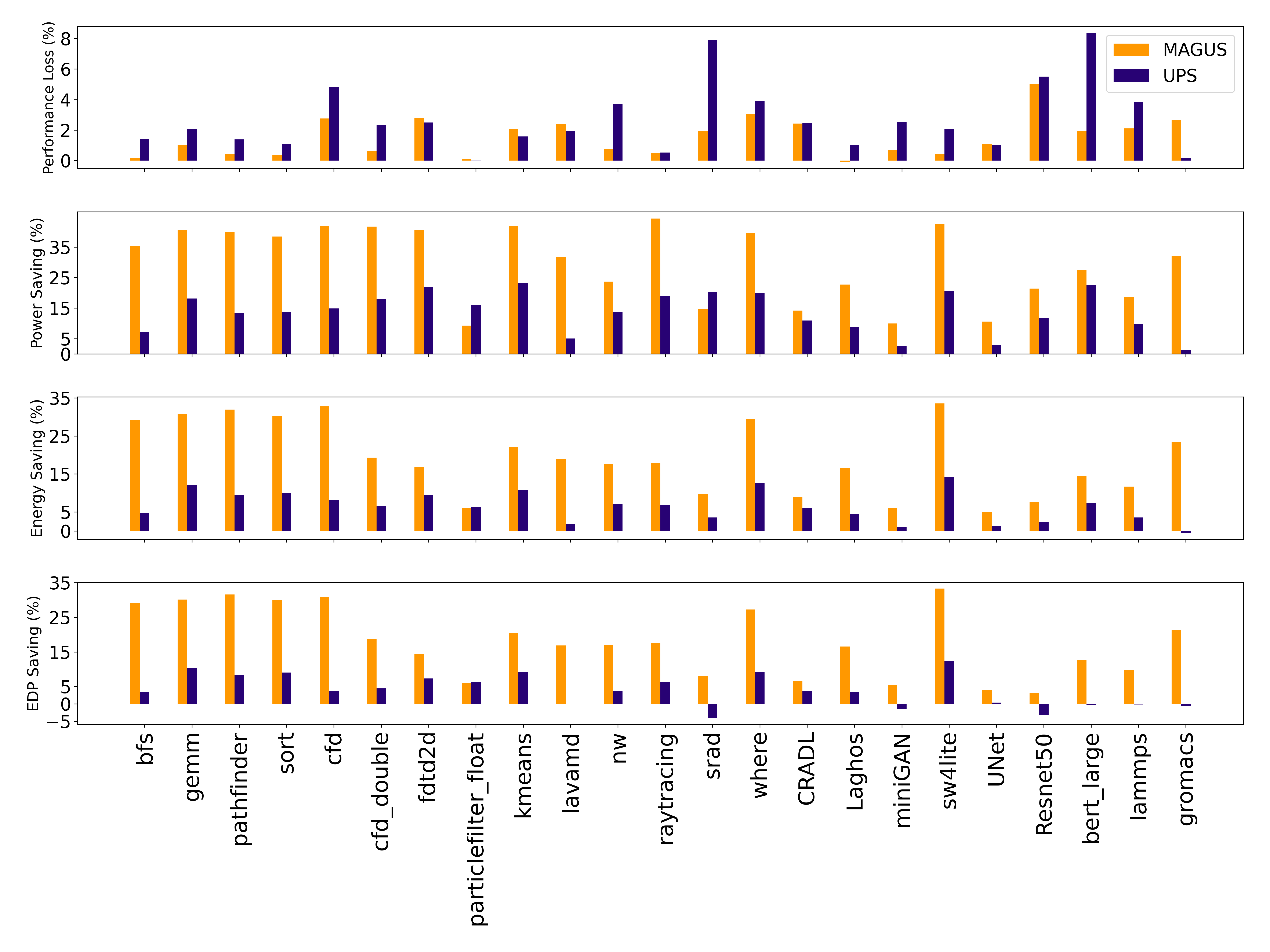}
    \caption{ Overall performance of the benchmarks and applications on Intel+A100. The X-axis lists the benchmarks and applications, while the Y-axis shows the corresponding metrics achieved by MAGUS and UPS against the baseline.}
    \label{fig:comparsion}
\end{figure*}

\subsection{Metrics}
We evaluate each method using four key metrics:
\begin{itemize}
    \item \textbf{Performance Loss}: Percentage increase in execution time compared to the baseline, measuring the runtime impact of uncore frequency scaling.
    \item \textbf{Package Power Saving}: Average reduction in CPU package power consumption relative to the baseline.
    \item \textbf{Energy Saving}: Total reduction in system energy consumption, including both CPU package and GPU (core and memory) energy, compared to the baseline.
    \item \textbf{Energy Delay Product (EDP)}: A composite metric that captures both energy efficiency and performance impact. Lower EDP values indicate better overall system efficiency, as they represent reduced energy consumption for a given performance level.
\end{itemize}

\section{Results} \label{result}

To account for performance variance and outliers when running the benchmarks and applications on real systems, each experiment was repeated at least ten times. Outliers were removed, and the average of the remaining results was calculated to ensure reliability. 

Our experiments evaluate uncore frequency scaling across single-GPU workloads on Intel+A100 and Intel+MAX1550, as well as multi-GPU workloads on Intel+4A100 systems. 

\subsection{Overall Performance} \label{overall_performance}
\begin{figure*}
    \centering
    \includegraphics[width=0.8\linewidth]{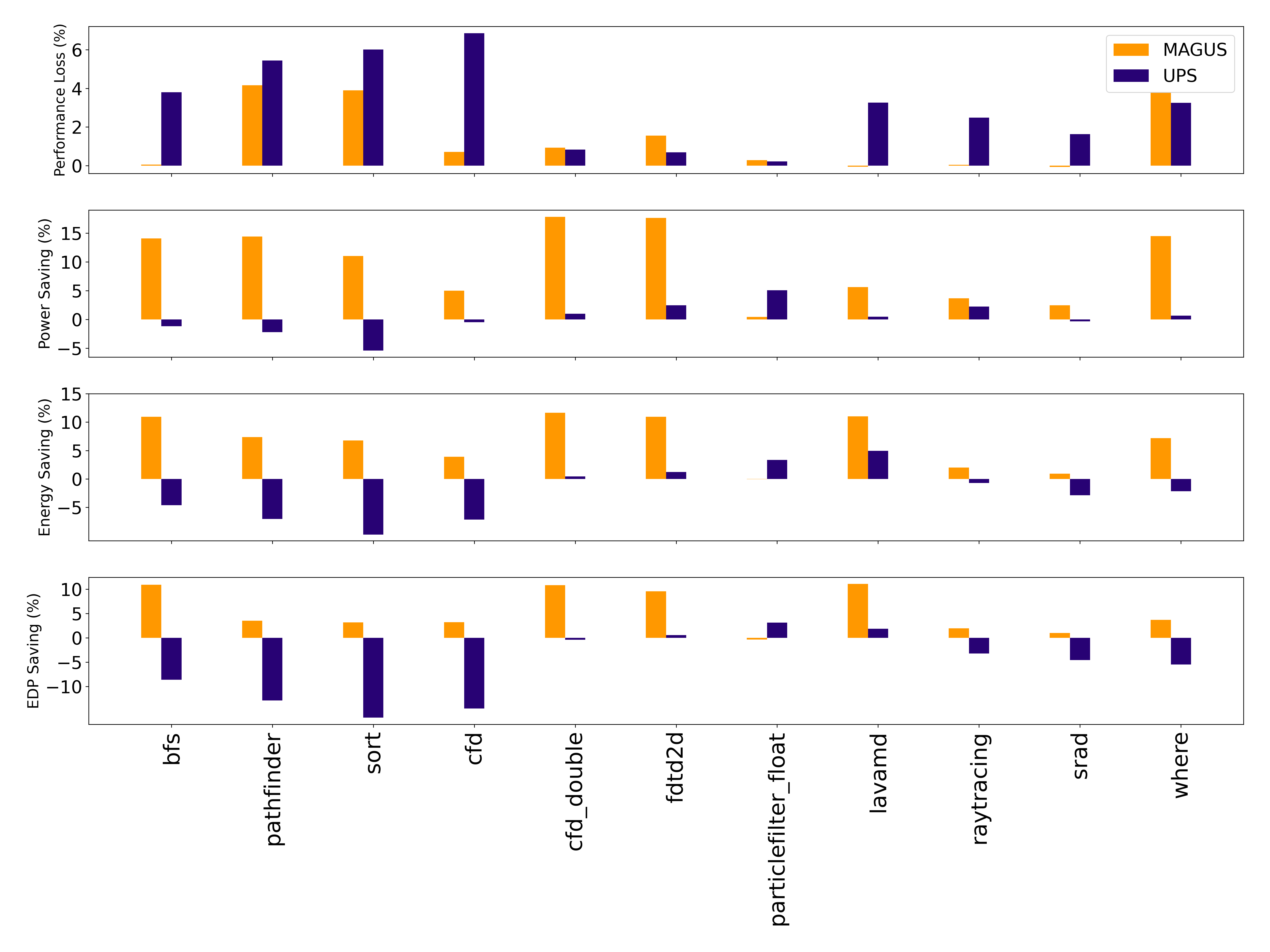}
    \caption{Overall performance on Intel+MAX1550. The X-axis lists the benchmarks, while the Y-axis shows the corresponding metrics achieved by MAGUS and UPS against the baseline.}
    \label{fig:pvc}
\end{figure*}

Figure \ref{fig:comparsion} compares different uncore frequency scaling methods  against the baseline in terms of performance loss, power savings, energy savings, and Energy Delay Product (EDP) savings on the Intel+A100 system, whereas Figure \ref{fig:pvc} presents the results on the Intel+Max1550 system.

As shown in the top plot of Figure \ref{fig:comparsion}, MAGUS consistently limits performance loss to below 5\%. For certain applications, such as sw4lite, gups, and nw, MAGUS results in no performance degradation. This is because these applications have low memory throughput requirements and are not CPU-intensive, allowing uncore frequency scaling without impacting overall performance. In the default uncore frequency scaling setting, as described in the previous section, the uncore frequency is reduced only when the CPU package power approaches the thermal design power (TDP). However, CPU package power rarely reaches TDP during GPU-enabled applications, as these workloads offload most of the computational tasks to the GPU, resulting in low CPU utilization. Despite this, GPU-enabled workloads still require high uncore frequency during periods of data movement and memory control operations. MAGUS continuously monitors memory throughput and dynamically adjusts the uncore frequency to minimize uncore power consumption. We observe that certain applications achieve higher CPU package power savings compared to others. This is because less CPU-intensive applications spend less time in high uncore frequency states, allowing for more frequent uncore downscaling and greater power efficiency. Reducing instantaneous power consumption helps prevent the aggregate power consumption of all applications from exceeding the system's total power budget, if one is in place. The energy savings in our experiments include both CPU package energy consumption and GPU power consumption. With MAGUS, all workloads achieve positive energy savings compared to the baseline, with savings of up to 27\%. By considering both the energy consumed and the execution time, EDP provides a more comprehensive evaluation of a system's efficiency compared to metrics that focus solely on energy or performance. A lower EDP value indicates higher energy efficiency. MAGUS achieves higher EDP savings up to 34\% compared to the UPS because the energy savings percentage achieved by MAGUS outweighs the percentage of performance loss. 

In Figure \ref{fig:pvc}, we make several observations regarding the Intel+Max1550 system. First, MAGUS maintains performance loss below 4\% while achieving up to 10\% energy savings and 10\% EDP savings, outperforming both the baseline and UPS. Second, for applications like fdtd2d and where, although MAGUS results in a higher performance loss than UPS, it achieves significantly greater energy savings up tp 10\%. This is because MAGUS applies a more aggressive uncore frequency tuning, reducing the frequency directly to the lower bound instead of gradually decreasing it. Third, UPS leads to negative energy savings for some applications due that it introduces a 7.9\% overhead in power consumption (see Section \ref{overhead analysis}). As a result, the power savings achieved by UPS are outweighed by the overhead it incurs.

Lastly, we observe both similarities and differences between the results on the two systems. MAGUS consistently achieves positive energy savings across all applications on both systems. However, for some applications on Intel+Max1550, UPS fails to achieve positive energy savings, unlike on Intel+A100. This discrepancy arises because UPS incurs a higher overhead in power consumption on Intel+Max1550 (7.9\%) compared to Intel+A100 (4.9\%) (see Section \ref{overhead analysis}).

\begin{table}[h]
    \centering
\caption{Overheads by MAGUS and UPS on different systems.}
\label{tab:magus_ups_comparison}
    \begin{tabular}{lcccc}
        \toprule
        System & \multicolumn{2}{c}{Power Overhead (\%)} & \multicolumn{2}{c}{Invocation Overhead (s)} \\
        \cmidrule(lr){2-3} \cmidrule(lr){4-5}
        & MAGUS & UPS & MAGUS & UPS \\
        \midrule
        Intel + A100   & 1.1\% & 4.9\% & 0.1s & 0.3s \\
        Intel + Max1550 & 1.16\% & 7.9\% & 0.1s & 0.31s \\
        \bottomrule
    \end{tabular}

\end{table}

\subsection{MAGUS Overheads} \label{overhead analysis}

A runtime inevitably introduces overheads. To quantify these overheads for MAGUS and UPS, we conduct an overhead analysis. We run MAGUS and UPS individually for 10 minutes without executing any applications, recording their power consumption and the time taken for each invocation (including hardware counter monitoring and phase detection). 
\emph{The power overhead} is calculated as the relative increase in power consumption introduced by each method. We also measure \emph{the invocation overhead}, in seconds, introduced by each method.

The results, shown in Table \ref{tab:magus_ups_comparison}, show that MAGUS incurs only ~1\% power overhead and requires just 0.1 seconds per invocation on both systems. In contrast, UPS introduces up to 7.9\% power overhead on Intel+Max1550 and takes approximately 0.3 seconds per invocation. MAGUS achieves 6× lower power overhead and 3× lower invocation overhead than UPS because it retrieves only a single hardware counter (memory throughput), whereas UPS accesses the MSR of each CPU core to read instructions retired and CPU cycles, in addition to DRAM power data.

\subsection{Detailed Analysis through a Case Study} \label{in-depth}

\begin{figure} 
    \centering
    \subfloat[]{%
       \includegraphics[width=1\linewidth]{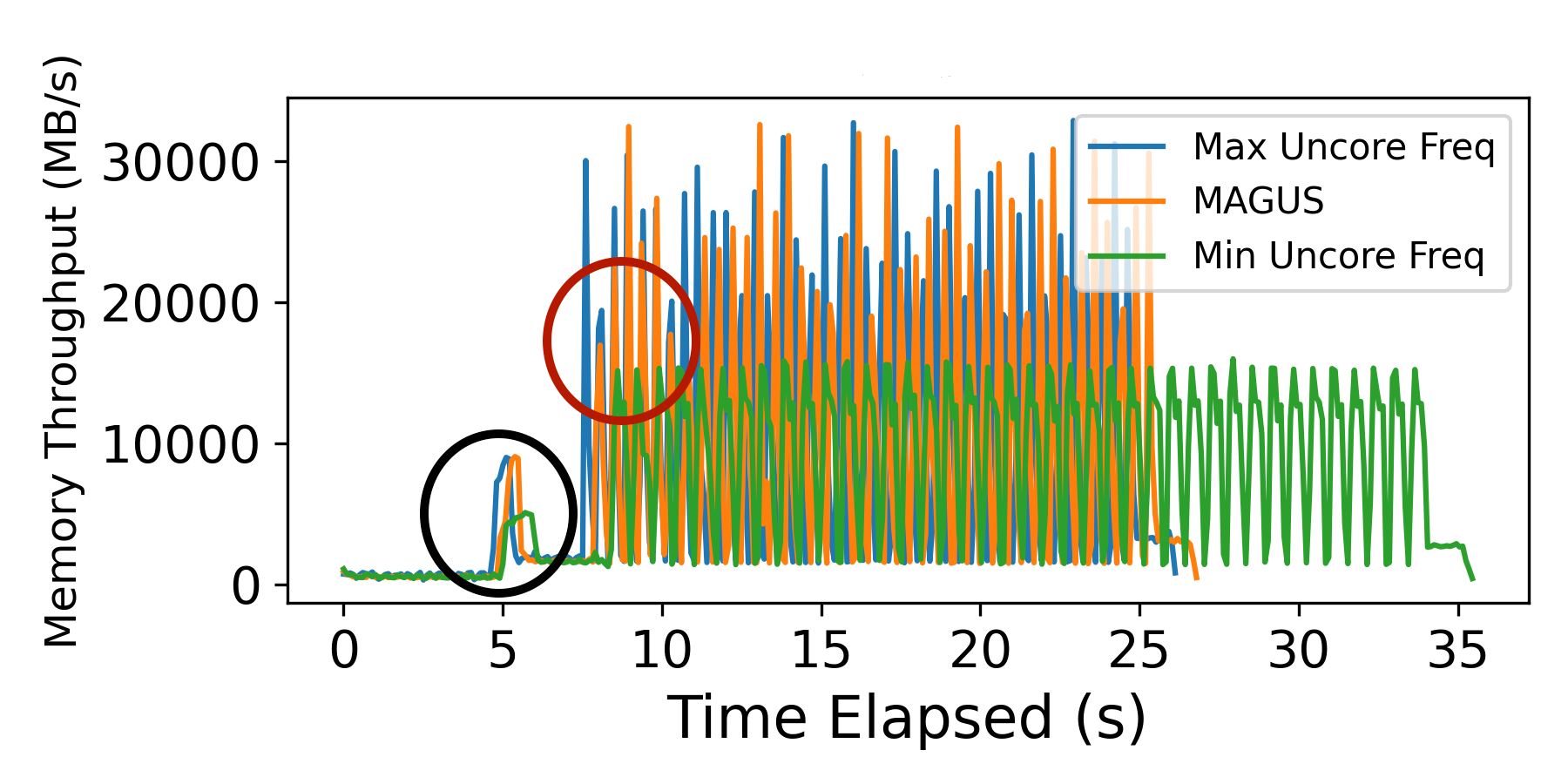}
       \label{fig:memory_throughput_srad_1}}\hfill
    \subfloat[]{%
        \includegraphics[width=1\linewidth]{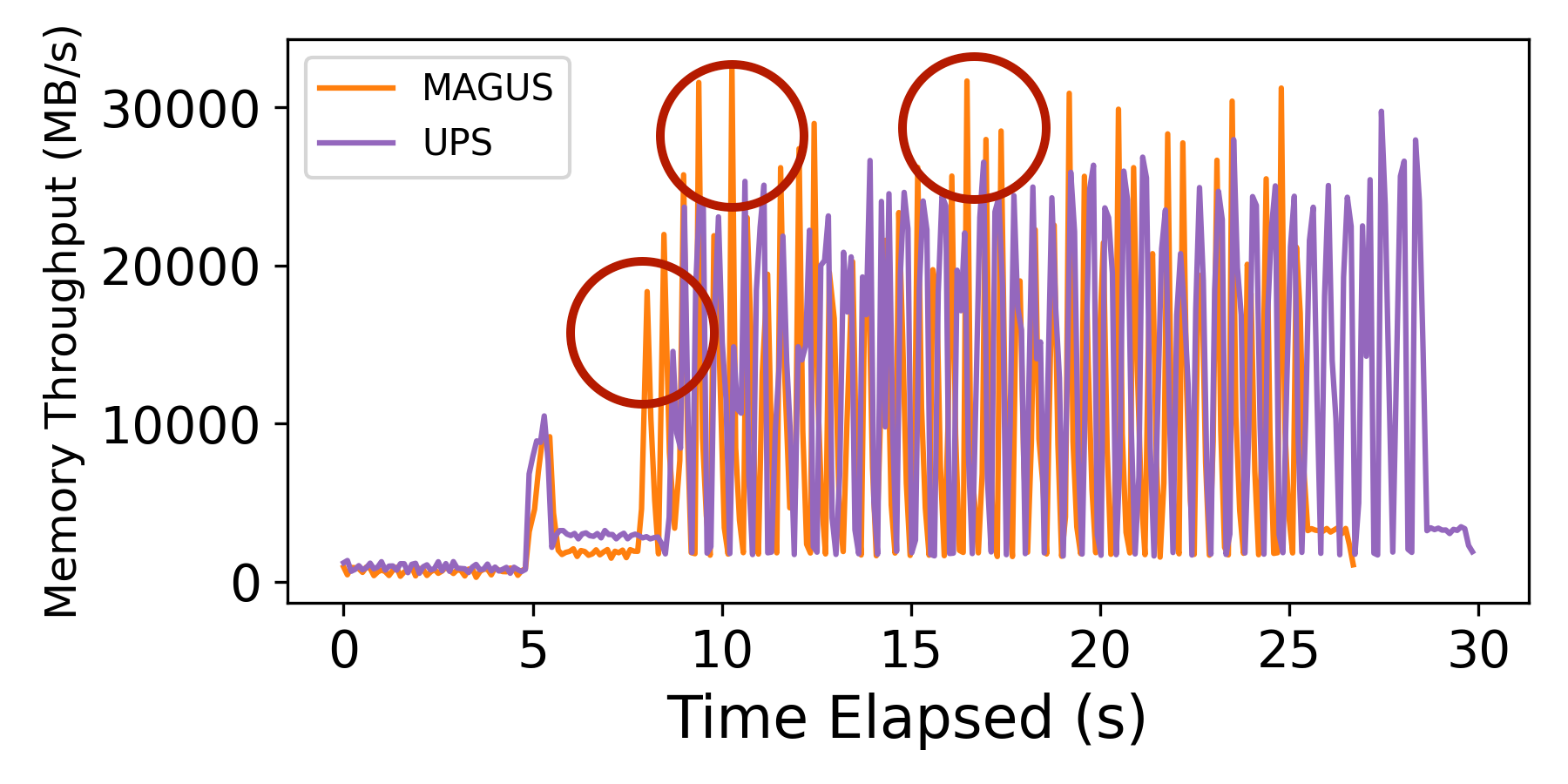}
        \label{fig:memory_throughput_srad_2}}
    \caption{Memory throughout of \emph{SRAD} under different uncore frequency settings. The top plot compares MAGUS with the minimum and maximum uncore frequency settings, while the bottom plot highlights the differences between MAGUS and UPS. The embedded circles highlight these differences.}
    \vspace{-10pt}
    \label{fig:memory throughput of srad}
\end{figure}

In this set of experiments, we conduct a zoom-in analysis to examine how different uncore frequency scaling methods behave using a case study of the \emph{SRAD} application on Intel+A100. This benchmark experiences high-frequency fluctuations in memory throughput, making it a particularly challenging case for uncore frequency tuning. By analyzing \emph{SRAD}, we gain deeper insights into how MAGUS dynamically adjusts uncore frequency compared to other methods.

In Figure \ref{fig:memory throughput of srad} (a), three memory throughput plots are presented for the \emph{SRAD} application under three scenarios: maximum uncore frequency (2.2 GHz), minimum uncore frequency (0.8 GHz), and MAGUS. Around the 5-second mark, the memory throughput under the minimum uncore frequency fails to match the level achieved by the maximum uncore frequency. In contrast, MAGUS successfully predicts changes in memory throughput trends, allowing it to reach comparable levels. Overall, MAGUS achieves memory throughput similar to the maximum uncore frequency while delivering an 8.68\% energy saving compared to Intel's default settings, with only a 3\% performance loss.

Next, we compare the memory throughput patterns of maximum uncore frequency, UPS, and MAGUS as shown in figure \ref{fig:memory throughput of srad} (b). We can observe that UPS fails to achieve the high memory throughput levels sustained by MAGUS (marked with read cycles). For workloads with frequent changes in memory throughput demand, response delays due to inherent hardware or software latencies often result in unmet throughput demands. To address this, MAGUS automatically detects phases with high-frequency changes in memory throughput and temporarily fixes the uncore frequency at its maximum level to avoid performance loss.

\begin{figure}
    \centering
    \includegraphics[width=1\linewidth]{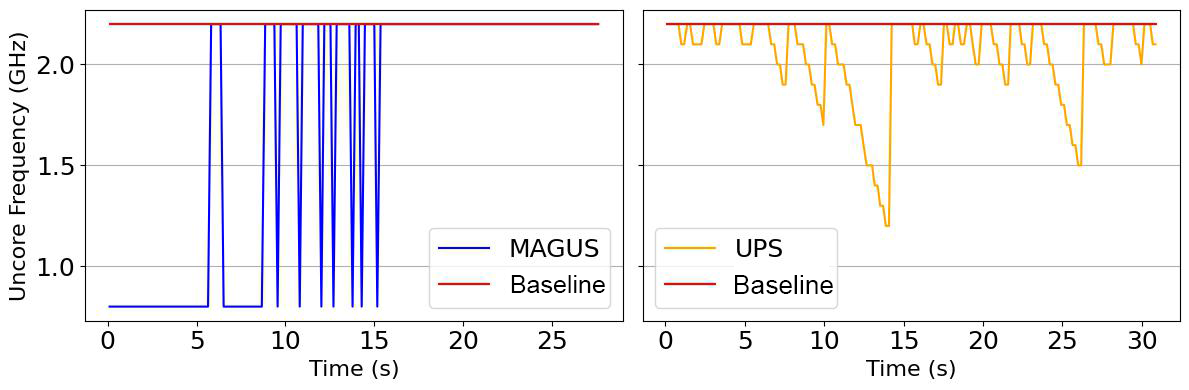}
    \caption{Uncore frequency of \emph{SRAD} under the baseline, UPS, and MAGUS.}
    \label{fig:srad_uncore_frequency}
\end{figure}

Figure \ref{fig:srad_uncore_frequency} illustrates how MAGUS leverages a high-frequency memory change detector to mitigate performance loss. MAGUS identifies high-frequency phases, such as between seconds 10 to 12.5 and after second 15, and locks the uncore frequency at its maximum (2.2 GHz) during these intervals to maintain performance stability. In contrast, UPS lacks this capability and continues to lower the uncore frequency after second 15, resulting in performance degradation. For the SRAD application, MAGUS achieves a 14\% reduction in CPU package power consumption compared to UPS's 20\%. However, MAGUS incurs only a 3\% slowdown, significantly lower than the 7.9\% slowdown observed with UPS. Consequently, MAGUS attains 8.68\% energy savings, outperforming UPS, which achieves only 3.5\% and results in negative EDP savings. By integrating high-frequency memory detection, MAGUS effectively minimizes performance loss when workloads exhibit frequent changes in memory throughput demand.

\subsection{Multi-GPU Analysis} \label{multi-gpu}

\begin{figure}
    \centering
    \includegraphics[width=0.8\linewidth]{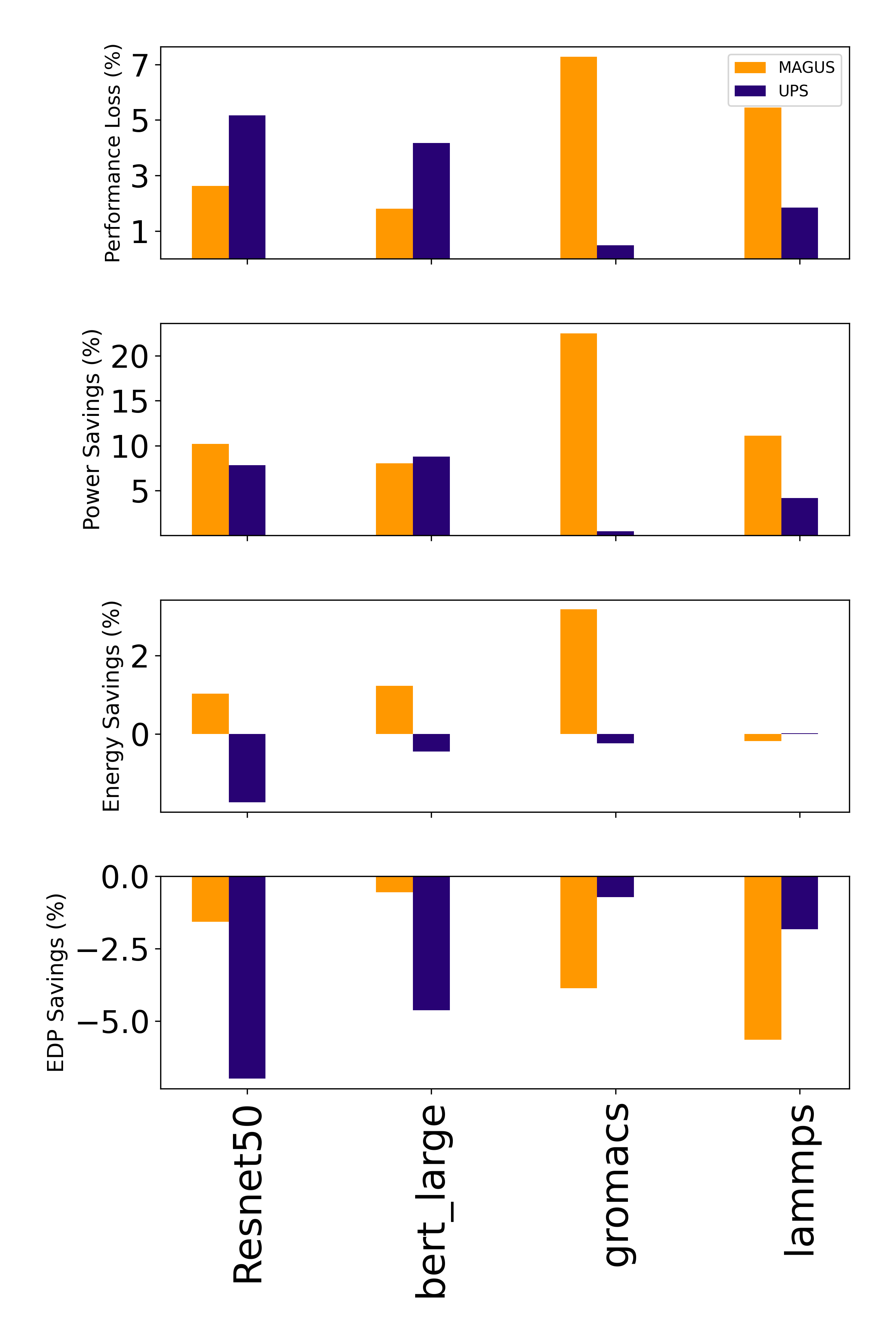}
    \caption{Multi-GPU results on Intel+4A100. The X-axis lists the benchmarks, while the Y-axis shows the corresponding metrics achieved by MAGUS and UPS against the baseline.}
    \label{fig:saving_multi_gpu}
    \vspace{-10pt}
\end{figure}

We extend our evaluation to multi-GPU scenarios using the Intel+4A100 system, focusing on AI-enabled applications and MLPerf benchmarks that effectively utilize multiple GPUs. Figure \ref{fig:saving_multi_gpu} presents the performance and energy efficiency results for these workloads.
We make several key observations. First, while MAGUS introduces a 7\% and 5.2\% performance overhead for GROMACS and LAMMPS respectively, it achieves around 21\% and 10\% CPU package power savings. Second, MAGUS provides greater or comparable energy savings than UPS for workloads such as ResNet50, BERT, and GROMACS. Third, unlike the single-GPU experiments, we see only modest energy savings  and occasionally negative EDP savings for both MAGUS and UPS. The primary reason is that \emph{idle power dominates on a multi-GPU system}. In this system,  the idle power for four A100-80GB GPUs is approximately 200W, significantly amplifying any runtime overhead compared to a single-GPU configuration, where a single A100-40GB GPU has an idle power of around 30W.
Because this idle power is \emph{a fixed operational cost} rather than a workload-specific component, including it in the total power calculation can obscure the actual efficiency gains achieved by uncore frequency scaling.

\begin{figure}
    \centering
    \includegraphics[width=0.8\linewidth]{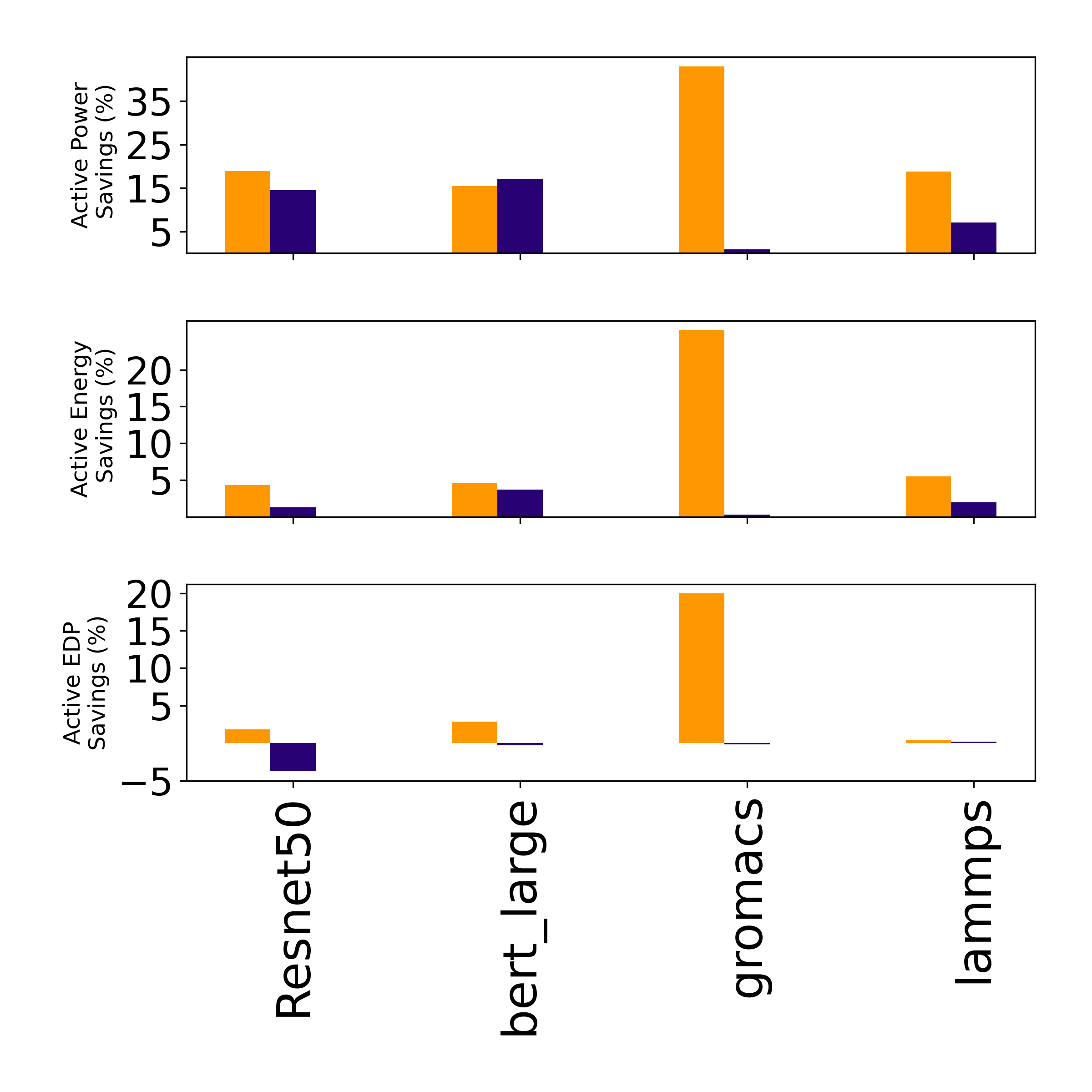}
    \caption{Active power and energy saving on Intel+4A100. Here, active power denotes the actual power consumption that a workload contributes.}
    \label{fig:multi-gpu-active-saving}
    \vspace{-5pt}
\end{figure}

To more accurately quantify MAGUS’s impact, we therefore measure \emph{active power and energy savings}, excluding idle power. For instance, if MAGUS reduces total power consumption from 200W to 150W on a system with 100W idle power, the active power saving is $\frac{\text{100} - \text{50}}{100}$ = 50\%. Figure\ref{fig:multi-gpu-active-saving} illustrates the active savings achieved through uncore frequency scaling, demonstrating that it can significantly reduce both active power and energy consumption by up to approximately 40\% for Lammps. Across all applications, MAGUS consistently achieves positive savings in terms of power savings, energy savings, and EDP. Furthermore, MAGUS outperforms both the baseline and UPS across all test cases. Specifically, MAGUS achieves approximately 5\% energy savings for ResNet50, BERT, and LAMMPS, while delivering up to 25\% energy savings with a 7\% performance loss for GROMACS.

\section{Related Work}

As large-scale high performance computing (HPC) systems evolve, power and energy efficiency have become critical priorities~\cite{DOE_goal}. 
Previous research has explored techniques such as CPU core Dynamic Voltage and Frequency Scaling (DVFS), CPU power capping, and memory DVFS to investigate the trade-offs between application performance and energy conservation \cite{ramesh2019understanding,walker2018hardware,wallace2016application,bailey2015finding,freeh2005using,ge2007cpu,hsu2005power,lim2006adaptive,rountree2007bounding, DRAM_FREQ, lefurgy2008power,petoumenos2015power,borghesi2015power}. 

Since the Sandy Bridge generation, Intel processors have enabled autonomous DVFS adjustments by modifying clock speeds and voltage dynamically, independent of software-specified settings \cite{intel-p-state,intel-p-state-2,dvfs}. Intel’s RAPL interface provides mechanisms for monitoring energy consumption and setting power limits in various CPU and DRAM domains\cite{david2010rapl,khan2018rapl}. Users can access RAPL data through model-specific registers (MSRs), \textit{sysfs} interface \cite{sysfs}, \textit{perf} \cite{perf} events, or the PAPI library \cite{weaver2012measuring}. 

With the increasing use of GPUs in HPC systems, power management has become an active research area. NVIDIA provides the NVIDIA Management Library (NVML), which enables users to control GPU power limits, core frequencies, and memory frequencies via the nvidia-smi interface \cite{NVML,nvidia-smi}. Significant efforts have been dedicated to modeling power and performance for GPU applications \cite{kandiah2021accelwattch,guerreiro2018gpgpu,arunkumar2019understanding,krzywaniak2020performance,guerreiro2019dvfs}. Research has also explored the impact of GPU core DVFS \cite{guerreiro2019dvfs,kraljic2022energy,fan2019predictable,ali2023performance} and GPU memory DVFS \cite{wang2020gpgpu} on energy efficiency in GPU workloads. 
Additionally, extensive studies are presented to optimize energy efficiency of deep learning training and inferences \cite{reguero2025energy,rajput2024benchmarking,panda2016conditional,8942147,you2023zeus}. 

Uncore frequency scaling has received relatively little attention in research and can be broadly classified into model-based and model-free approaches. In model-based approaches \cite{sundriyal2018core, zhang2024fcufs}, analytical or machine learning models predict optimal uncore frequencies by monitoring multiple hardware counters in real-time. For instance, Sundriyal et al. \cite{sundriyal2018core} develop power and performance models to adjust the uncore frequency for optimal power efficiency during application execution. Zhang et al. \cite{zhang2024fcufs} train a neural network to predict application performance and power consumption, leveraging multi-objective optimization to minimize power usage while limiting performance loss.
In contrast, model-free approaches \cite{gholkar2019uncore, guermouche2022combining} bypass complex models by dynamically detecting phase transitions between compute-intensive and memory-intensive regions to guide uncore frequency scaling. UPScavenger \cite{gholkar2019uncore} is a pioneering model-free solution that monitors multiple hardware counters to adjust uncore frequency based on workload phases. 

\section{Conclusion}\label{Conclusion}
As HPC systems increasingly adopt heterogeneous architectures, balancing power efficiency and performance is critical. Our work highlights uncore frequency scaling as a previously underexplored yet impactful strategy for energy optimization in GPU-accelerated environments. MAGUS addresses this by combining lightweight memory throughput monitoring with dynamic phase detection, achieving up to 27\% energy savings and 26\% EDP reduction while keeping performance loss under 5\%. Moreover, MAGUS only introduces ~1\% overhead in power consumption. Its success in both single- and multi-GPU environments underscores the need for specialized power management strategies beyond traditional CPU-centric approaches.

\bibliographystyle{ACM-Reference-Format}
\bibliography{bib/hpdc}

\end{document}